\documentclass[aps,prc,amsmath,twocolumn,showpacs,superscriptaddress,
amssymb,showkeys]{revtex4}
\usepackage{graphicx}
\usepackage{epsfig}
\usepackage{enumerate}
\usepackage{verbatim}
\usepackage{color}

\begin{document}

\title{Testing the continuum discretized coupled channel method for deuteron induced reactions}

\author{N. J. Upadhyay}
\affiliation{National Superconducting Cyclotron Laboratory, Michigan
State University, East Lansing, MI 48824-1321, USA}
% ----------------------------------------------------------------------
\author{A. ~Deltuva}
\affiliation{Centro de F\'{\i}sica Nuclear da Universidade de Lisboa,
P-1649-003 Lisboa, Portugal}
% ----------------------------------------------------------------------
\author{F. M. Nunes}
\affiliation{National Superconducting Cyclotron Laboratory, Michigan
State University, East Lansing, MI 48824-1321, USA}
\affiliation{Department of Physics and Astronomy, Michigan State
University, East Lansing, MI 48824-1321, USA}
% ----------------------------------------------------------------------

\date{\today}

\begin{abstract}
The Continuum Discretized Coupled Channels (CDCC) method is
a well established theory for direct nuclear reactions which includes
breakup to all orders. Alternatively, the 3-body problem can be
solved exactly within the Faddeev formalism which explicitly
includes breakup and transfer channels to all orders. With the aim
to understand how CDCC compares with the exact 3-body Faddeev
formulation, we study deuteron induced reactions on: i) $^{10}$Be at
$E_{\rm d}= 21.4, 40.9 \; {\rm and } \; 71$ MeV; ii) $^{12}$C at $E_{\rm d} = 12 \; {\rm and } \; 56$ MeV; and
iii) $^{48}$Ca at $E_{\rm d} = 56$ MeV.
We calculate elastic, transfer and breakup cross sections.
Overall, the discrepancies found for elastic scattering are small with
the exception of very backward angles. For transfer cross sections at low energy $\sim$10 MeV/u, CDCC
is in good agreement with the Faddeev-type results and the discrepancy increases with beam energy.
On the contrary, breakup observables obtained with CDCC are in good agreement with Faddeev-type results
for all but the lower energies considered here.
\end{abstract}



\maketitle

%%%%%%%%%%%%%%%%%%%%%%%%%%%% INTRODUCTION %%%%%%%%%%%%%%%%%%%%%%%%%%%%%%
\section{Introduction}

The construction and upgrade of various radioactive beam facilities worldwide
continue to open up new avenues for the study of direct nuclear
reactions involving rare isotopes \cite{review1,review2}.
At the low energy regime (around 10-20 MeV/u) one can explore the structure of the
exotic nucleus in detail. In this context, deuteron induced reactions in inverse kinematics
are of great  interest because of the reduced Coulomb barrier, when compared to heavier
targets. These include, elastic, inelastic, transfer and breakup reactions.
Single neutron transfer using (p,d) or (d,p) is also becoming a common tool to explore
neutron capture for astrophysics (e.g. \cite{c14dp}).

Given the long history of using deuteron induced reactions (in normal kinematics) for
studying nuclear structure, a number of reaction theories are well established. With the understanding
that breakup strongly affects the reaction mechanism when the loosely bound deuteron is involved,
theories including deuteron breakup were developed \cite{jonsop,jontan,cdcc}. However, until
recently the accuracy of the methods has not been quantified. Because the standard reaction methods
are not based on any normal ordering perturbation theory, it is not possible to estimate the accuracy
intrinsically. One relies on an independent exact formulation of the problem.

Faddeev devised a method which  couples rearrangement, elastic and breakup channels \cite{fad} by expanding
the three-body wavefunction into three components. For any given three-body Hamiltonian, the Faddeev
method provides the exact solution.
For few-nucleon reactions an integral equation formulation
for the transition operators  \cite{ags,gloc}
proposed by Alt, Grassberger, and Sandhas (AGS)
is commonly used in the momentum-space framework.
 One critical issue in this method is
the well known Coulomb problem. Screening and renormalization techniques \cite{alt:80a,deltuva05} have been developed
to handle Coulomb, which opened the path for applications to deuteron induced nuclear reactions \cite{nunes2,deltuva09}.
This method can now be used to benchmark existing nuclear reaction theories.

One method of practical use in (d,p) and (p,d) reactions, is the finite range adiabatic method (ADWA) \cite{jontan} which takes
into account deuteron breakup to all orders in the transfer channel while making an important simplification of the continuum.
ADWA has been studied in detail in \cite{nguyen10} and compared to exact Faddeev predictions in \cite{nunes11a,nunes11b}.
Those studies demonstrate the validity of ADWA for (d,p) reactions at low energies ($E\approx 10$ MeV/u), populating states
with low angular momentum and small separation energy. Results in \cite{nunes11b} indicate that the
accuracy of ADWA deteriorates with increasing beam energy,
angular momentum and separation energy of the final state.

ADWA is the precursor of the Continuum Discretized Coupled Channel Method (CDCC) \cite{cdcc}, which does not make
approximations in the treatment of the breakup states. While ADWA provides a method to compute transfer alone,
CDCC can be used for all, elastic, breakup and transfer. CDCC is now a standard tool to study
deuteron induced breakup reactions \cite{scholar}. In CDCC, the 3-body scattering problem is solved through a set of coupled
Schr\"{o}dinger-like differential equations. The breakup is included to all orders by expanding the full three-body wave function in terms of a complete basis of the projectile's bound and continuum states. A discretization by averaging over energy is common to handle the coupling between the continuum states.
In addition to elastic and breakup, attempts have been made to study transfer reactions using the CDCC wavefunction \cite{nunes1,ogata,jepp,kee,nunes2}. In these methods, the exact three-body wavefunction in the transfer amplitude
is replaced by the CDCC wavefunction. Thus, starting from one single three-body Hamiltonian,
CDCC can provide, in a consistent manner, elastic, breakup and transfer cross sections.

In theory, CDCC should provide an exact solution to the three-body problem. In practice,
due to the truncation of the model space this may not be true. For example, it is easy to demonstrate
that the truncated deuteron CDCC wavefunction does not provide the correct asymptotic form for a bound state in a rearrangement channel.
A first attempt to compare CDCC and Faddeev was performed in Ref. \cite{nunes2}.
Good agreement was found for elastic scattering and breakup for deuterons on $^{12}$C and $^{58}$Ni while
disagreement was observed for $^{11}$Be on protons at E = 38 MeV/u in both breakup and transfer
observables. The reaction calculations for $^{12}$C and $^{58}$Ni did not include bound states in the rearrangement channels and therefore one may naively expect the CDCC solution to provide the correct answer to the problem.
The disagreement found for the single reaction where a bound transfer state was included, questioned the reliability of CDCC and called for a more systematic study.

The aim of this work is to quantify the accuracy of CDCC
in computing elastic, breakup and transfer cross sections and establish a range of validity.
This will be done by direct comparison with exact Faddeev-AGS calculations (FAGS). Our test cases include
deuteron scattering from
 $^{10}$Be, $^{12}$C,  and
 $^{48}$Ca. These reactions are studied as a function of beam energy, chosen to match available experimental data.

In Section II, we describe the details of the CDCC and AGS formalisms. In Section III
we present the numerical details of the calculations, followed by the results and discussions, in Section IV.
Finally, the summary and conclusions are drawn in Section V.

% ----------------------------------------------------------------------

%%%%%%%%%%%%%%%%%%%%%%%%%%%%% FORMALISM %%%%%%%%%%%%%%%%%%%%%%%%%%%%%%%%
\section{Formalism}
\label{theory}

For describing the $A(d,d)A$,  $A(d,p)B$, and  $A(d,pn)A$ reactions we consider an effective 3-body problem $p + n + A$ ,
such that B is the ground state $B =n + A$. The Hamiltonian is then written as
\begin{equation}
\label{hamil}
H_{3b}\,=\,\hat{T_R}\,+\,\hat{T_r}\,+\,U_{\rm pA}\,+\,U_{\rm nA}\,+\,
V_{\rm pn}
\end{equation}
where the kinetic energy operators, $\hat{T_R}$ and $\hat{T_r}$ can be expressed
in any of the three Jacobi coordinate pairs \cite{book}. The deuteron is modeled
by a real potential, $V_{\rm pn}$, reproducing its binding energy and the $np$ low
energy phase shifts.
The p-A and n-A interactions are phenomenological optical potentials,
$U_{\rm pA}$ and $U_{\rm nA}$, respectively. They include absorption
from all the channels which are otherwise not included in the model
space.

\subsection{The CDCC Method}%%%%%%%%%%%%%%%%%%%%%%%%%%%%%%%%%%%%%%%%%%%

In the CDCC approach, using the above mentioned 3-body Hamiltonian
(Eq. \ref{hamil}), the 3-body wave equation of the total system is written as
\begin{equation}
\label{schro}
\left[\,H_{3b}\,-\,E\,\right]\,\Psi^{\rm CDCC}({\bf r}, \bf{R})\,=\,0 \;,
\end{equation}
where ${\bf r}$ and ${\bf R}$ are the Jacobi coordinates for the p-n and
d-A systems, respectively. Here $E$ is the total energy of the system
in the center of mass (c.m.) frame.

Keeping in mind that the deuteron is a loosely bound system, we consider
deuteron breakup states explicitly. The eigenstates of the deuteron,
$\phi({\bf r})$ satisfy the eigenvalue equation $(T_r+V_{np})\,
\phi({\bf r})\,=\,\epsilon\,\phi({\bf r})$.
To reduce the computational cost, in this work we ignore the spins of
the particles. We perform the standard partial wave decomposition in terms of the orbital angular momentum of neutron relative
to the proton ($l$)  and the discretization of the deuteron continuum (indexed i) by slicing the continuum into bins and averaging over momentum \cite{book,cdcc,scholar}. In this manner, the projectile is represented by:
\begin{equation}
\label{binwave}
\phi_{i}({\bf r})\,=\,\sum_l\,i^l\,{\frac{u_{il}(r)}{r}}\, Y_l(\hat{\bf r}) \;,
\end{equation}
where  $i=0$ refers to the deuteron ground state and $i>0$ to the $np$ continuum bin $i$
corresponding to momentum $k_i=\sqrt{{2\,\mu_{pn}\,\epsilon_i}/{\hbar^2}}$ with $\mu_{pn}$ being the reduced mass of proton and
neutron. The projectile continuum is truncated by restricting the orbital momentum ($l \le l_{max}$) and
the energy of the $np$ breakup bins  ($\epsilon_i \le \epsilon_{max}$). The limits are chosen such that the observables of interest are converged.

A partial wave decomposition in terms of the relative angular momentum between projectile and target ($L$)
is also performed. Thus, the full 3-body CDCC wave function can be written as:
\begin{equation}
\label{wave}
\Psi^{\rm CDCC}({\bf r}, {\bf R})\,=\sum_{\alpha}\,i^{l+L} {\frac{u_{il}(r)}{r}}\,
\chi_{\alpha}(R)\, Y_{l_{\alpha}}(\hat{\bf r})\,Y_{L_{\alpha}}(\hat{\bf R}) \;,
\end{equation}
where we use $\alpha$ as a general index representing the quantum numbers $\alpha=\{ilL\}$.

Substituting Eq.(\ref{wave}) in Eq.(\ref{schro}), a set of coupled-channel equations is
obtained for $\chi_{\alpha}(R)$,
\begin{eqnarray}
 \left [   -\frac{\hbar^2}{2 \mu_{dA}}  \left ( \frac{d^2 }{ d R^2} - \frac{L(L{+}1)} {R^2} \right )
   + V_{\alpha \alpha}(R)  + \epsilon_i - E \right ]
 \chi_ {\alpha}(R)  \nonumber \\
 \hspace{2cm} +  \sum _ {\alpha'\ne\alpha}
 i ^ {L' - L} ~ V_{\alpha \alpha'}(R)  \chi_{\alpha'}  (R)= 0\ ,
\label{cdcceq}
\end{eqnarray}
where $\mu_{dA}$ is the reduced mass of $d+A$, and $E_{\alpha}=E- \epsilon_i$.
Eq.(\ref{cdcceq}) is solved with the scattering boundary conditions for large $R$:
\begin{equation}
\chi_{\alpha}(R) \rightarrow i/2 \left[ H_{\alpha}^-(KR) \delta_{\alpha \alpha_i} - H_{\alpha}^+(KR) S_{\alpha \alpha_i}\right] \;.
\end{equation}
Here, $H^-$ and $H^+$ are the Coulomb Hankel function \cite{book} and $S_{\alpha \alpha_i}$ is the S-matrix.
The coupling potentials in Eq.(\ref{cdcceq}) are defined by:
\begin{equation}
V_{\alpha\alpha^\prime}(R) = \langle \phi_{il}(r) | U_{\rm pA} + U_{\rm nA} |\phi_{i^\prime l^\prime}(r)\rangle,
\end{equation}
and contain both nuclear and Coulomb parts, which are expanded in terms of multipoles $Q$ up to $Q_{max}$.

As mentioned before, the CDCC wavefunction Eq.(\ref{wave}) can be used in
the exact transition amplitude for $A(d,p)B$.  The post form for the transition amplitude is then:
\begin{equation}
T\,=\,\langle\,\chi^{(-)}_{\rm pB}\,\phi_{\rm nA}\,|\,
V_{\rm pn}\,+\,U_{\rm pA}\,-\,U_{\rm pB}\,|\,\Psi^{CDCC}\,\rangle,
\label{tmatrix}
\end{equation}
where $\chi^{(-)}_{\rm pB}$ is the distorted wave generated by the auxiliary potential $U_{\rm pB}$
(depending on {$\bf R^\prime$}, the p-B relative coordinate), and $\phi_{\rm nA}$ is the bound
state for the final n-A system.
The difference $(U_{\rm pA} - U_{\rm pB})$ is called the {\it remnant
term}, which is sometimes neglected for convenience.
It should be stressed that, in the case of exotic nuclei, it can be very significant.
In our study it is always included.

\subsection{The Faddeev-AGS method}%%%%%%%%%%%%%%%%%%%%%%%%%%%%%%%%%%%%%

The Faddeev formalism explicitly includes  the
deuteron $d+A$, the proton $(nA)+p$, and the neutron $(pA)+n$ components.
Three sets of Jacobi variables are used such that
each component is treated in its proper basis.
In the AGS approach the three particle scattering is described in terms
of the transition operators $T_{\beta \alpha}$, where $\alpha$ ($\beta$)
corresponds to the initial (final) channel in the odd-man-out notation.
The transition operators obey the AGS equations  \cite{ags} that are
a system of coupled integral equations
\begin{equation}
\label{agseq}
T_{\beta \alpha} = \bar{\delta}_{\beta \alpha}\,G_0^{-1}+
\sum_{\gamma=1}^3\,\bar{\delta}_{\beta \gamma}\,t_{\gamma}\,
G_0 T_{\gamma \alpha}.
\end{equation}
Here $\bar{\delta}_{\beta \alpha} = 1 - \delta_{\beta \alpha}$ and
$G_0 = (E+i0 - H_0)^{-1}$ is the free resolvent
  with $E$ being the
total energy in the three-body c.m. system. The two-body transition
operator for each interacting pair is derived from
the pair potential $v_{\gamma}$ via the Lippmann-Schwinger equation
\begin{equation}
t_{\gamma}= v_{\gamma}\,+\,v_{\gamma}\,G_0\,t_{\gamma}.
\end{equation}
The scattering amplitude $X_{\beta \alpha}$ is
the on-shell matrix element of  $T_{\beta \alpha}$ calculated between
initial and final channel states as
\begin{equation}
X_{\beta \alpha} = \langle \phi_{\beta} | T_{\beta \alpha}
| \phi_{\alpha} \rangle.
\end{equation}
In the case of elastic scattering $\beta = \alpha$,
while  $\beta = 0$ for breakup and $0 \ne \beta \ne \alpha$
for transfer reactions.
The AGS equations (\ref{agseq}) are solved in momentum space using
partial wave decomposition. The proton-nucleus
Coulomb interaction is treated using the method of screening and
renormalization \cite{deltuva05,alt:80a}. Further numerical details
can be found in Refs.~\cite{delt1,delt2,delt3,delt4}.

% ----------------------------------------------------------------------

%%%%%%%%%%%%%%%% NUMERICAL DETAILS AND INPUTS %%%%%%%%%%%%%%%%%%%%%%%%%%
\section{Numerical Details and Inputs}

In this section we present the pair interactions necessary to define the three-body Hamiltonian
as well as some details for the model space used. Both CDCC and Faddeev-AGS are computationally expensive,
thus, without sacrificing the final goal, we make simplifications to the $p+n+A$ system.
To reduce the number of channels, we have neglected the spins of the particles and thus all the spin-orbit terms
in the interactions are not included. Also, a single gaussian interaction is used to describe the deuteron
and its continuum \cite{gaus}
\begin{equation}
V(r)\,=\,-\,V_0\,e^{-\,(r/r_0)^2},
\end{equation}
with $V_0$ = 72.15 MeV and $r_0$ = 1.484 fm. The potential reproduces the binding energy of deuteron
and low-energy p-n $\,{}^3S_1$ phase shifts
 but fails heavily in other partial waves. While the comparison of CDCC and Faddeev-AGS is meaningful given that the same
Hamiltonian is used in both formulations, a direct comparison of the predicted cross sections with data should be avoided.
We have chosen for all our test cases energies at which data exists and thus this study retains its relevance to experiments.

An important ingredient to the three-body Hamiltonian, the nucleon-nucleus optical potentials, are all taken
from the global parameterization CH89 \cite{ch89}. The optical potentials are energy dependent and should
be considered within the context of the reaction model.
In CDCC calculations, $U_{\rm pA}$ and $U_{\rm nA}$ are calculated at half the deuteron incident energy $E_d$/2, while
the auxiliary potential, $U_{\rm pB}$, is calculated at the proton
energy $E_{\rm p}$ corresponding to the exit channel. The Faddeev-AGS calculations with exactly
the same choice of $U_{\rm pA}$ and $U_{\rm nA}$ as in CDCC yield
results for elastic and breakup observables only as there is no transfer channel
with complex $U_{\rm nA}$. We therefore perform also the  Faddeev-AGS calculations
where a real neutron-nucleus potential is used in the partial wave corresponding to the final bound state.
The same  interaction is used to produce
the neutron-nucleus bound state $\phi_{nA}$ in Eq.(\ref{tmatrix}). We take a Woods-Saxon central potential with
standard geometry: radius $r_0=1.25$ fm and diffuseness $a=0.65$ fm. The depths are adjusted to reproduce the neutron
separation energy of the relevant state. Details are given in Table \ref{tab-bs}.

%@@@@@@@@@@@@@@@@@@@@@@@@@@@@@@@@ Table 1 @@@@@@@@@@@@@@@@@@@@@@@@@
\begin{table}[h!]
\vspace{10pt}
\centerline{
\begin{tabular}{ccccccc}
\hline
\hline
\vspace{0.1cm}
Nucleus (A) && $nl$  && S$_{\rm n}$ (MeV) &&  $V_{\rm nA}$ (MeV)
\\
\hline
$^{10}$Be && 2$s$ && 0.504 && 57.064
\\
&
\\
$^{12}$C && 1$p$ && 4.947 && 39.547
\\
\\
$^{48}$Ca && 2$p$ && 5.146 && 48.905
\\\hline
\hline
\end{tabular}}
\caption{\label{tab-bs}Parameters of the n-A binding potential used
in the CDCC calculations. The Coulomb radius and the diffuseness
are taken to be $r_0$ = 1.25 fm and $a_0$ = 0.65 fm, respectively.
S$_n$ is the neutron separation energy, $V_{\rm nA}$ is the depth of
the n-A binding potential and $nl$ are the quantum numbers
describing the ground state of the nucleus B=A+n.}
\end{table}

For both CDCC and Faddeev-AGS, convergence needs to be studied in detail.
While elastic and transfer observables offered a minor challenge from the point of view of convergence,
the same cannot be said of breakup distributions. Even though spins were neglected
to reduce computational cost, breakup calculations  proved to be at the limit of our numerical
capabilities. Due to technical difficulties in the Faddeev-AGS, breakup calculations were performed without Coulomb.
In Section \ref{cdcc-model-space} we present the CDCC model space based on transfer observables. For those cases in which
the model space was insufficient to ensure convergence of the breakup cross sections,
the predictions are presented with bands instead of lines, estimated on the
convergence behavior with increasing $l$. In Section \ref{fags-model-space} we present the Faddeev-AGS model space
chosen to ensure convergence of all calculated observables.

\subsection{CDCC Model space} %%%%%%%%%%%%%%%%%%%%%%%%%%%%%%%%%
\label{cdcc-model-space}

The $^{10}$Be(d,p)$^{11}$Be reaction is studied at three beam
energies E$_{\rm d} = 21.4, 40.9 \mbox{ and } 71$ MeV.
The calculations include p-n partial waves up to $l_{max}=4$ and
$Q_{max}=4$ multipoles in the expansion of the coupling potentials.
The coupled equations are integrated up to $R_{max}$ = 60 fm with total
angular momentum $J_{max} = 40$ for E$_{\rm d} = 21.4$ MeV
and $J_{max} = 50$ for E$_{\rm d} =40.9 \mbox{ and } 71$ MeV.

Calculations for $^{12}$C(d,p)$^{13}$C reaction includes p-n
partial waves $l \le 6$ and coupling potentials expanded up to
$Q_{max}= 6$. The coupled equations are integrated up to
$R_{max}=60$ fm with $J_{max} = 30$ for  E$_{\rm d} = 12$ MeV and
$J_{max}= 50$ for E$_{\rm d} = 56$ MeV.

Calculations for $^{48}$Ca(d,p)$^{49}$Ca reaction includes p-n
partial waves $l \le 6$ and coupling potentials expanded up to
$Q_{max} = 6$. The coupled equations are integrated up to
$R_{max} = 80$ fm with $J_{max} = 70$ for  E$_{\rm d} = 56$ MeV.

\begin{figure}[ht!]
\begin{center}
\includegraphics[scale=0.35]{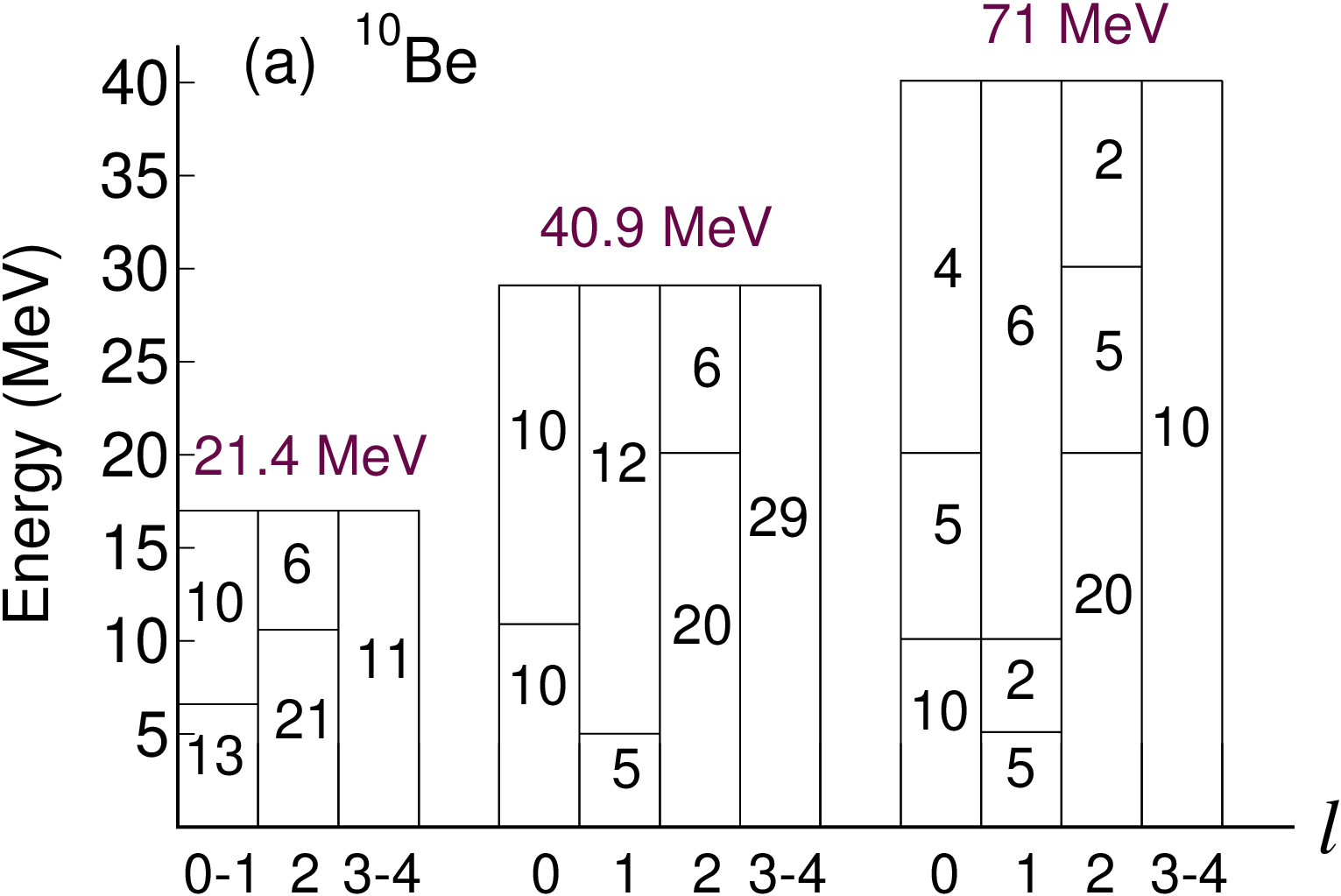}

\vspace{0.4cm}
\includegraphics[scale=0.35]{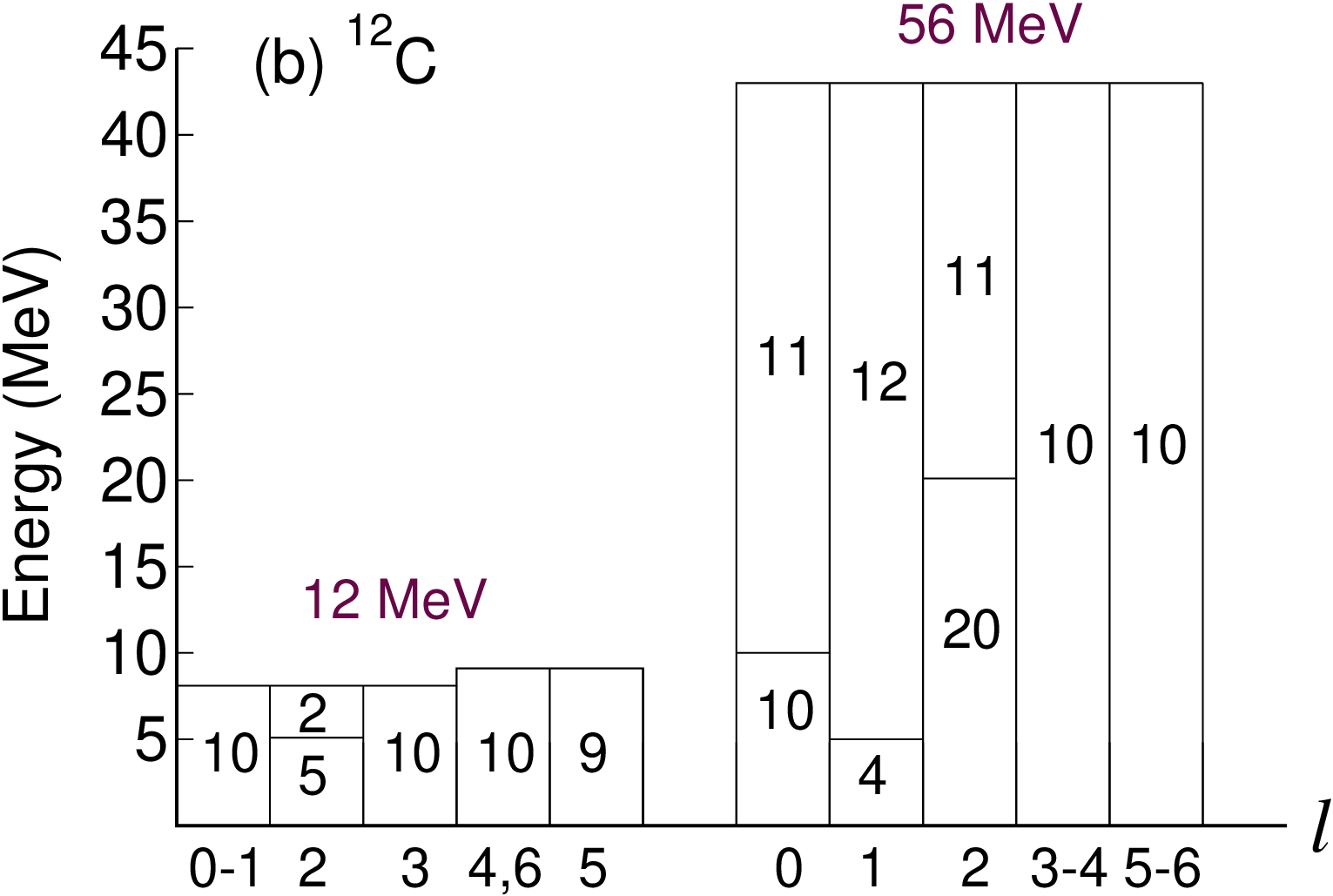}

\vspace{0.4cm}
\includegraphics[scale=0.35]{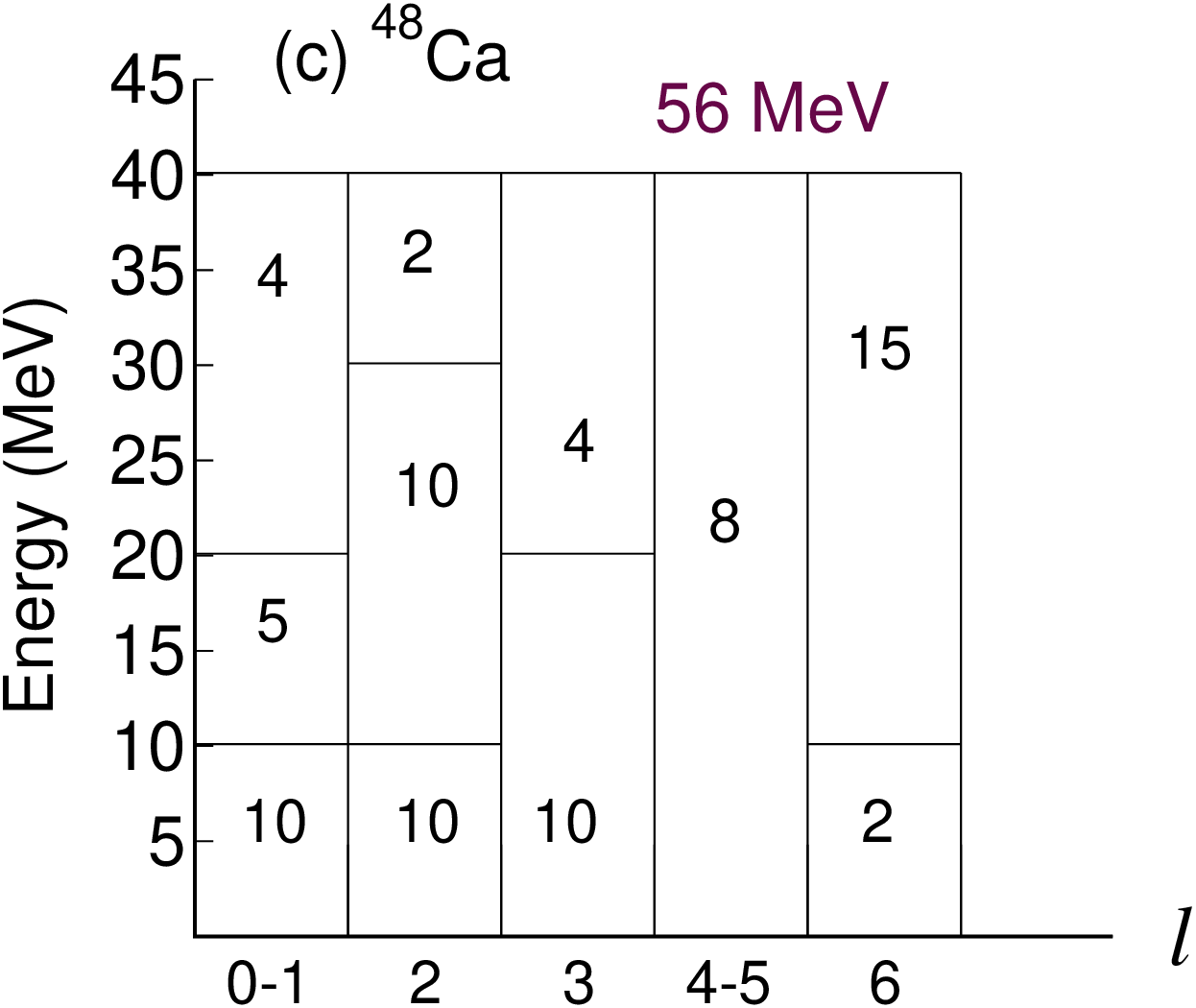}
\end{center}

\vspace{-1em}
\caption{\label{bins}Schematic detail of deuteron continuum
used in the calculations, showing number of bins and the
energy range as a function of $l$: (a) d+$^{10}$Be, (b) d+$^{12}$C
and (c) d+$^{48}$Ca.}
\end{figure}

For all calculations, the radial integration was performed in steps of $0.06$ fm and the
momentum average defining the bins was performed on a mesh with $20$ points.
The schematic detail of the bin structure used for the discretized
deuteron continuum in these calculations, is shown in Fig. \ref{bins}.
%&&&&&&&&&&&&&&&&&&&&& Figure 1 &&&&&&&&&&&&&&&&&&&&&&&
%&&&&&&&&&&&&&&&&&&&&& Ends &&&&&&&&&&&&&&&&&&&&&&&&&&&

When considering transfer, in addition to the above mentioned
model space, the non-locality of the transfer kernel needs to be considered.
The details of the non-local parameters used are
presented in Table \ref{tab-param}. Due to the strong repulsive couplings at short
distances, for $^{10}$Be(d,p)$^{11}$Be  at E$_{\rm d} = 71$ MeV, it
was necessary to introduce an L-dependent internal cut in the couplings.
%($cutl = -3.5$ as defined in \cite{fresco}).

%@@@@@@@@@@@@@@@@@@@@@@@@@@@@@@@@ Table 2 @@@@@@@@@@@@@@@@@@@@@@@@@
\begin{table}[h!]
\vspace{10pt}
\centerline{
\begin{tabular}{l r r r}
\hline
\hline
\vspace{0.1cm}
target & $E_d$ (MeV) & Centration (fm) &  Non-local width (fm) \\
\hline
 	    & 21.4  & 2.16 & 14 \\
$^{10}$Be & 40.9  & 1.92 & 20 \\
 	    & 71  & 1.32 & 12 \\
\hline
$^{12}$C & 12 & -0.96 & 6 \\
 	& 56 & -0.66 & 7 \\
\hline
$^{48}$Ca &  56 & -0.78 & 7 \\
\hline
\hline
\end{tabular}}
\caption{\label{tab-param}Parameters of the non-local transfer kernel used
in the CDCC calculations (details can be found in \cite{fresco}).}
\end{table}
%@@@@@@@@@@@@@@@@@@@@@@@@@@@@@@@@@@ Ends @@@@@@@@@@@@@@@@@@@@@@@@@@

\subsection{Faddeev-AGS Model space} %%%%%%%%%%%%%%%%%%%%%%%%%%%%%%%%%
\label{fags-model-space}

The AGS equations are solved using partial wave decomposition.
However, in contrast to CDCC, three  bases corresponding to the
three choices of Jacobi momenta are used.
The maximum we include for the p-n partial waves is $l_{max}=5$.
For the $^{10}$Be target, n-$^{10}$Be  partial waves up to $l_{max}=8$ and
p-$^{10}$Be  partial waves up to $l_{max}=16$ are included.
For the $^{12}$C target, n-$^{12}$C  partial waves up to $l_{max}=8$ and
p-$^{12}$C  partial waves up to $l_{max}=18$ are included.
For the $^{48}$Ca target, n-$^{48}$Ca  partial waves up to $l_{max}=10$ and
p-$^{48}$Ca  partial waves up to $l_{max}=20$ are included.
The maximal total angular momenta considered is  $J_{max} = 60$.
Depending on reaction and energy, some of these cutoffs can
be reduced without affecting the quality of the results.
The integrals are discretized using Gaussian quadrature rules with
about 50 grid points for each Jacobi momentum.
The Coulomb screening radius is $R_C = 16$ fm.
The results are well converged except for
the breakup cross sections at small angles.
In that case a special treatment of the pure Coulomb breakup
term as described in Ref.~\cite{nunes2} improves the convergence.
However, even with this special treatment including
$R_C=50$ fm and proton-nucleus partial waves up to $l_{max}=40$
we were unable to get better than 20\% accuracy for small angles.
We therefore present breakup results neglecting the Coulomb force,
since the main goal of the present work is the comparison
of the methods and not the analysis of experimental data.
The Faddeev-AGS calculations without Coulomb force are well converged;
in this case  $l_{max}=10$ for p-A partial waves is sufficient.

% ----------------------------------------------------------------------

%%%%%%%%%%%%%%%%%%%%%% RESULTS AND DISCUSSION  %%%%%%%%%%%%%%%%%%%%%%%%%
\section{Results and Discussion}
\label{results}

In this section we present results for elastic scattering, transfer and breakup.
For many of the chosen cases, data is available. However, because we neglect spin, the
direct comparison with data is not appropriate. It is the comparison between
the methods that is meaningful.

%@@@@@@@@@@@@@@@@@@@@@@@@@@@@@@@@ Table 3 @@@@@@@@@@@@@@@@@@@@@@@@@
\begin{table}[h!]
\vspace{10pt}
\centerline{
\begin{tabular}{l c c r}
\hline
\hline
\vspace{0.1cm}
Label & $U_{\rm pA}$  & $U_{\rm nA}$ &  nA-bound
\\
\hline
FAGS & $E_d$/2 & $E_d$/2 & no
\\
FAGS1 & $E_d$/2 & $E_d$/2 & yes
\\
FAGS2 & $E_p$ & $E_d$/2 & yes
\\
\hline
\hline
\end{tabular}}
\caption{\label{tab-fadd}Types of Faddeev-AGS calculations being performed, the labels used, the energies at which the associated interactions were determined and whether a neutron-nucleus potential supports a bound state.}
\end{table}

While the reference calculations are performed with the neutron and proton optical potentials calculated at
half the deuteron energy, there are small useful variations in the interactions that should be considered carefully.
For clarity we describe the various labels that will appear in the following subsections in Table \ref{tab-fadd}.
For elastic and breakup, the standard CDCC and Faddeev-AGS calculations are labeled CDCC and FAGS, respectively.
These take the nucleon optical potentials $U_{pA}$ and $U_{nA}$ at half the deuteron energy in all partial waves
and therefore produce no bound states in the nucleon-nucleus subsystems.

For there to be a transfer channel in Faddeev-AGS, the relevant interaction needs to hold a bound state.
The only difference between FAGS and FAGS1 is the inclusion of a neutron-nucleus bound state (by taking a real interaction
for the nA subsystem in the partial wave for which a bound state exists). Comparing FAGS and FAGS1
for elastic and breakup reaction tells
us about the importance of coupling to the transfer channel.

In evaluating transfer matrix elements, the calculations labeled CDCC take the nucleon-nucleus interactions ($U_{pA},U_{nA}$)
in the initial state to be at half the deuteron energy $E_d/2$ and the auxiliary
interaction $U_{pB}$ to be at the proton energy in the exit channel $E_p$, closely mimicking the physical process.
Therefore it makes sense to consider also the Faddeev-AGS counterpart FAGS2, where $U_{pA}$ is determined at $E_p$.
We note that FADD2 and FADD of Refs.~\cite{nunes11a,nunes11b} correspond to our FAGS1 and FAGS2, respectively.
For completeness we also explore the effect of this subtle difference in the initial Hamiltonian in CDCC,
by performing CDCC2 (where all proton optical potentials  are calculated at $E_p$, including the one in the incoming channel).

%%%%%%%%%%%%%%%%%%%%%% ELASTIC  %%%%%%%%%%%%%%%%%%%%%%%%%

\subsection{\label{es}Elastic Scattering}

The simplest reaction  to study is elastic scattering. The elastic differential cross section is
sensitive to the asymptotic behavior of the three-body wavefunction in a very particular region
of phase space, namely when the neutron and proton are bound. It is thus one basic test for a complete
reaction theory.

\begin{figure}[t!]
\begin{center}
\includegraphics[scale=0.33]{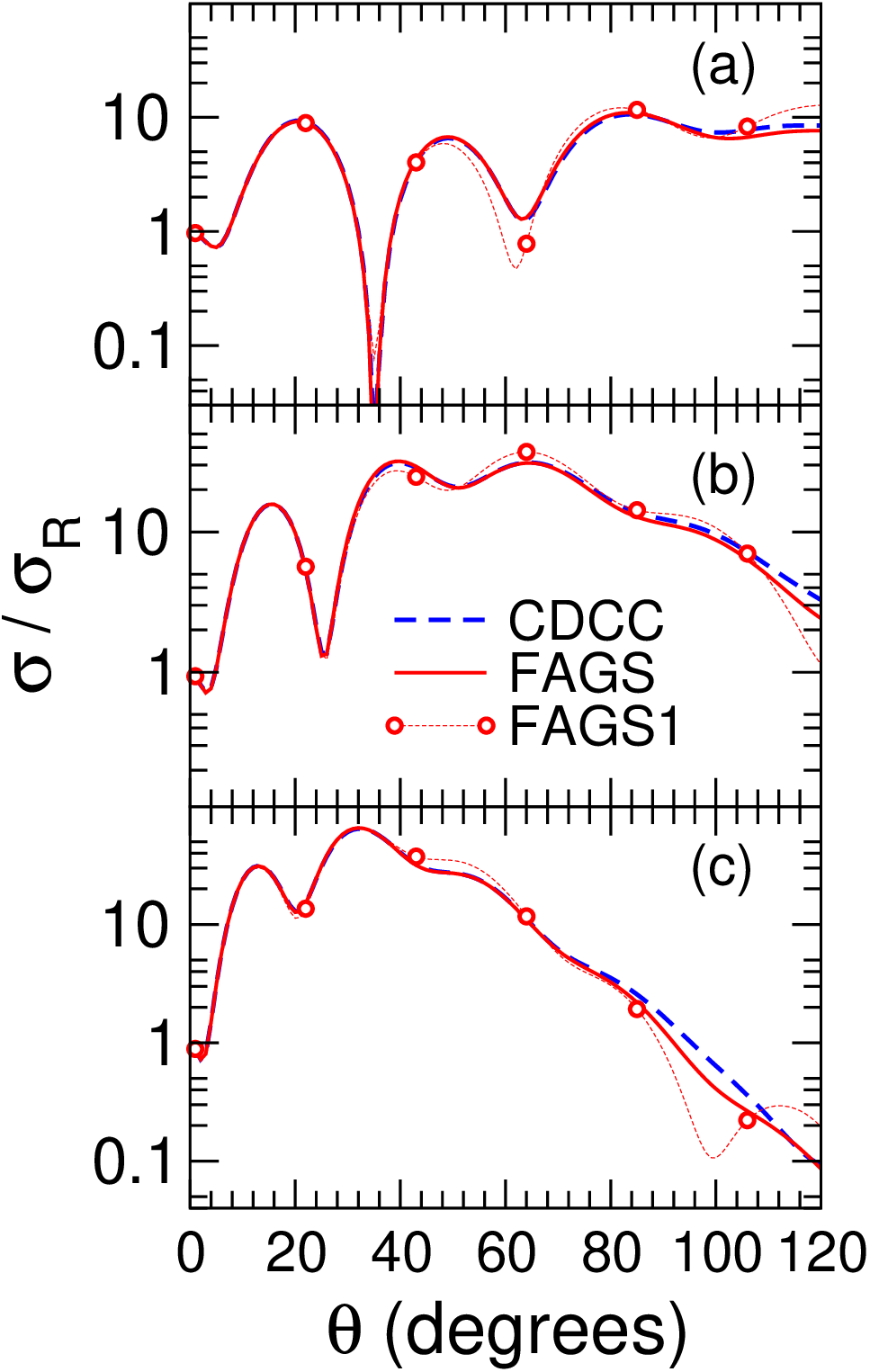}
\end{center}

\vspace{-1em}
\caption{\label{be10el}Elastic cross section for d$+^{10}$Be: (a) $E_{\rm d}$ = 21.4 MeV,
(b)$E_{\rm d}$ = 40.9 MeV, and (c) $E_{\rm d}$ =
71 MeV.}
\begin{center}
\includegraphics[scale=0.33]{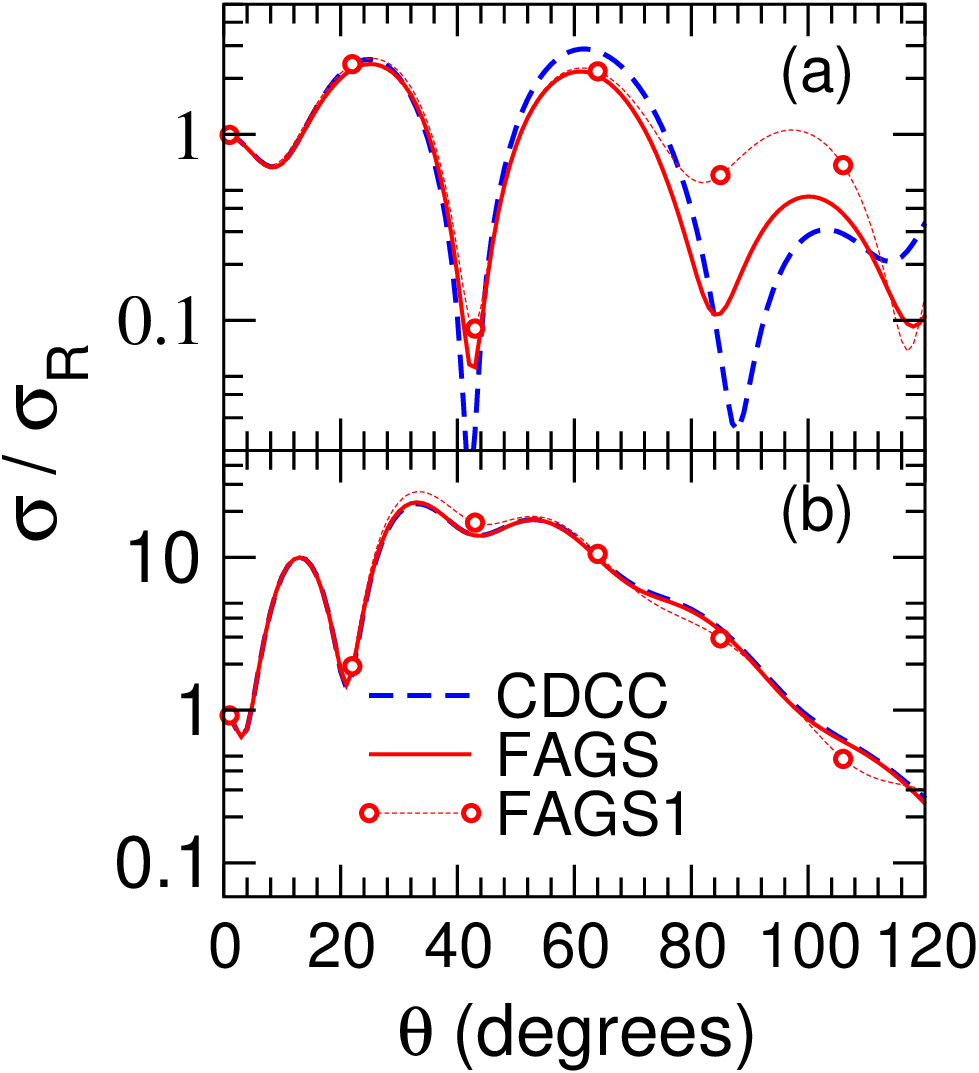}
\end{center}

\vspace{-1em}
\caption{\label{c12el}Elastic cross section for
d$+^{12}$C: (a) $E_{\rm d}$ = 12 MeV
and (b) $E_{\rm d}$ = 56 MeV.}
\begin{center}
\includegraphics[scale=0.33]{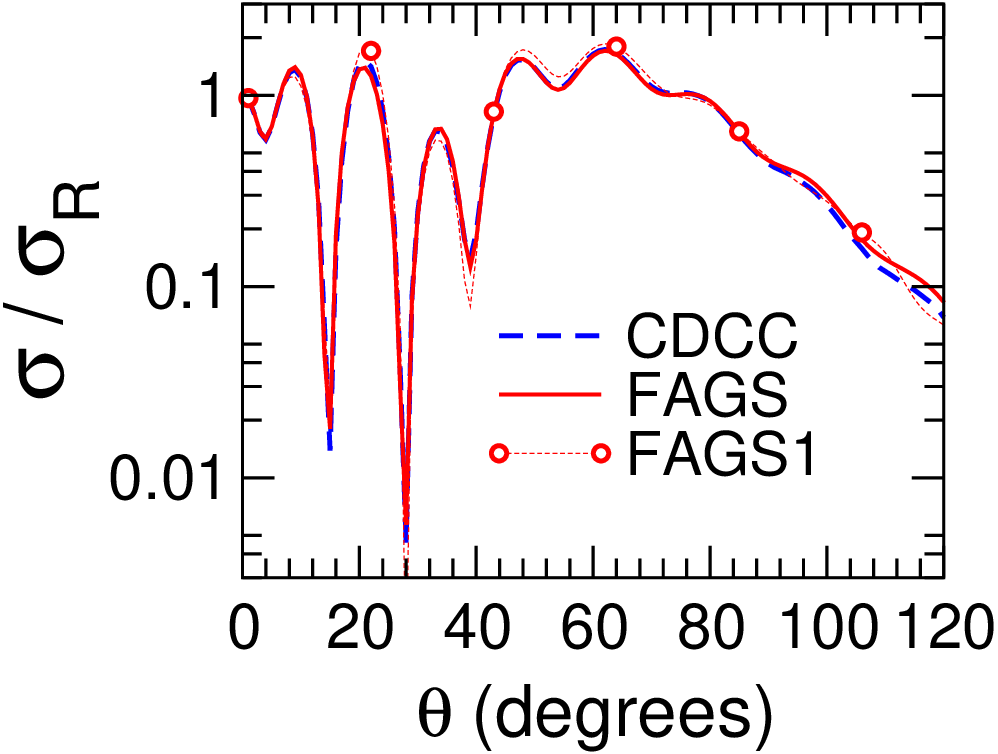}
\end{center}

\vspace{-1em}
\caption{\label{ca48el}Elastic cross section for
d$+^{48}$Ca at $E_{\rm d}$ = 56 MeV.}
\end{figure}

The elastic cross sections for deuterons on $^{10}$Be, $^{12}$C and $^{48}$Ca are shown in Figs. \ref{be10el}, \ref{c12el} and \ref{ca48el}, respectively. Plotted are the standard CDCC (dashed) and FAGS (solid) angular distributions, as well as the results of FAGS1 (circles-dotted), when the transfer channel is included. The differential cross section is divided by the corresponding
Rutherford cross section to allow for inspection at larger angles. Overall
CDCC and FAGS agree, with exception of d$+^{12}$C at 12 MeV, where some discrepancy is found beyond $40^{\circ}$. The inclusion of the transfer couplings has a minor effect, as can be seen by comparing FAGS and FAGS1, which produces only small modifications in the distributions, mainly at large angles.

%%%%%%%%%%%%%%%%%%%%%% TRANSFER  %%%%%%%%%%%%%%%%%%%%%%%%%%%%%%%%%%%%%%%%%%%%%%%%%

\subsection{\label{trans}Transfer Reaction}

\begin{figure}[t!]
\begin{center}
\includegraphics[scale=0.315]{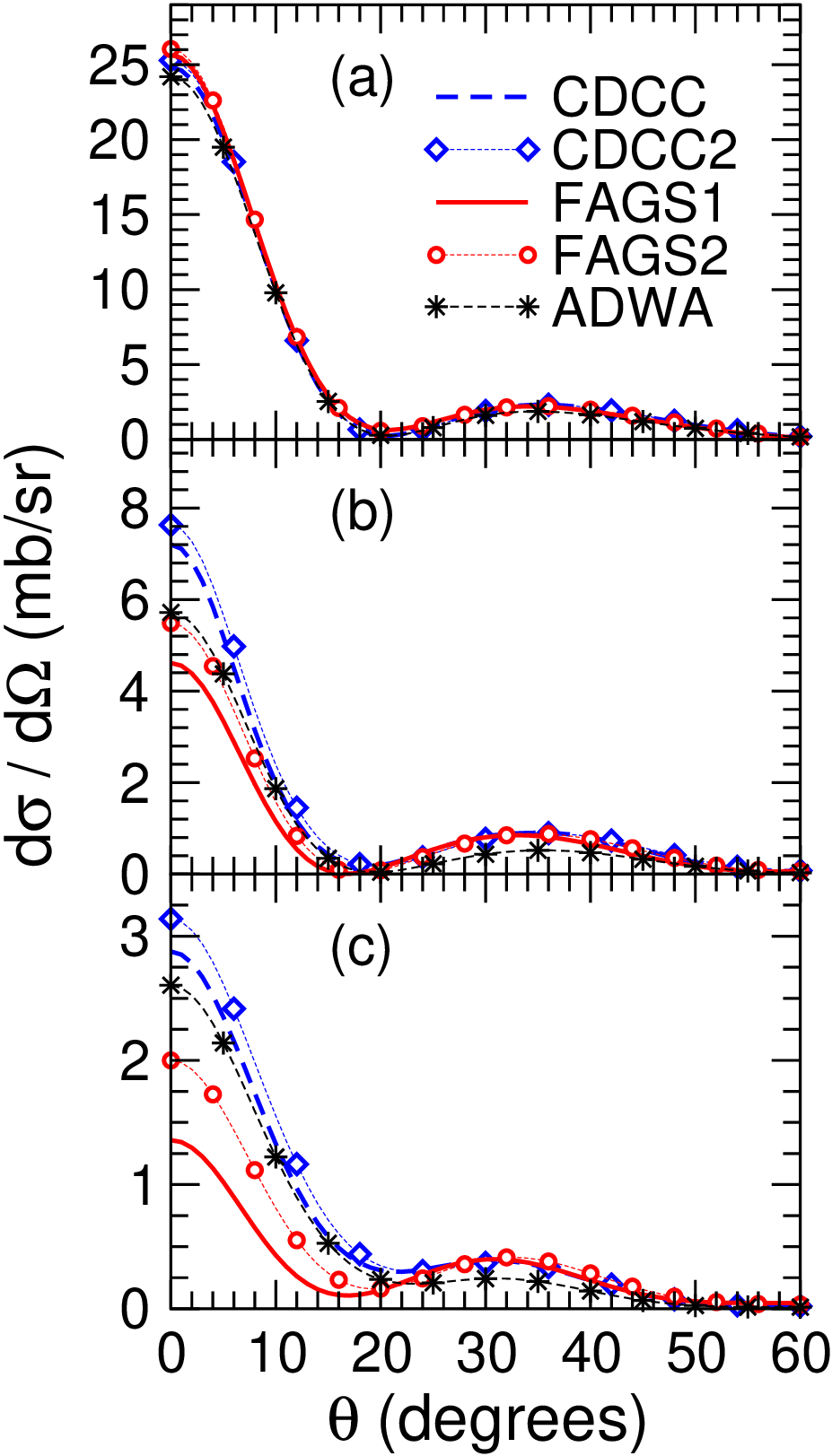}
\end{center}

\vspace{-1.2em}
\caption{\label{be10tr}Angular distribution for $^{10}$Be (d, p) $^{11}$Be:
(a) $E_{\rm d}$ = 21.4 MeV, (b) $E_{\rm d}$ = 40.9 MeV, and (c) $E_{\rm d}$ = 71 MeV.}
\begin{center}
\includegraphics[scale=0.315]{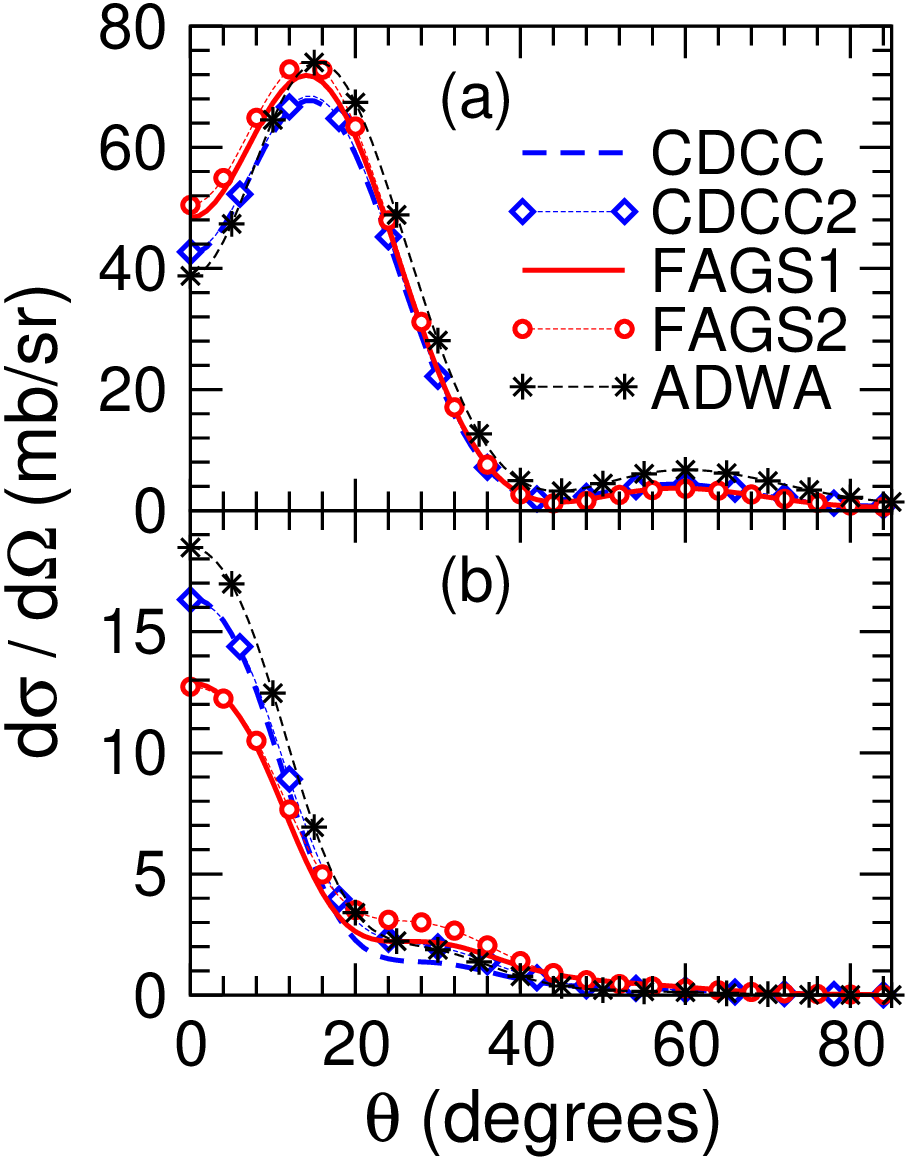}
\end{center}

\vspace{-1.2em}
\caption{\label{c12tr}Angular distribution for $^{12}$C (d, p) $^{13}$C:
(a) $E_{\rm d}$ = 12 MeV and (b) $E_{\rm d}$ = 56 MeV.}
\begin{center}
\includegraphics[scale=0.315]{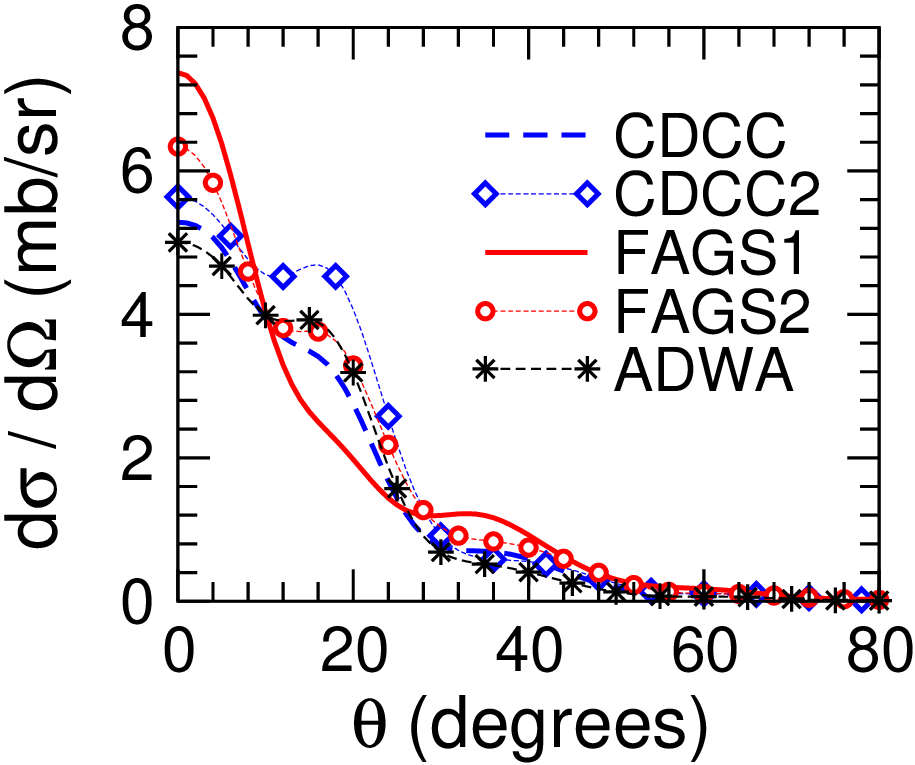}
\end{center}

\vspace{-1.2em}
\caption{\label{ca48tr}Angular distribution for $^{48}$Ca (d, p) $^{49}$Ca at $E_{\rm d}$ = 56 MeV.}
\vspace{-1.2em}
\end{figure}
Contrary to elastic scattering, the transfer process is typically sensitive to the three-body scattering
wavefunction at the surface of the target, and therefore poses different challenges for reaction theory.
Here, all cases refer to transfer to the ground state (g.s.) of the final nucleus.

In Fig. \ref{be10tr}, we present the angular distribution
for $^{10}$Be (d, p) $^{11}$Be reaction calculated at $E_{\rm d}$ = 21.4
MeV, 40.9 MeV and 71 MeV, respectively.
We first compare the standard CDCC (dashed) and standard FAGS1 (solid).
We observe that at $E_{\rm d}$ = 21.4 MeV, the two methods are in perfect
agreement, but significant discrepancies appear at higher beam energy.
To determine how much of those discrepancies can be attributed to the choice of the energy
at which the optical potentials are evaluated, it is useful to consider the auxiliary calculations CDCC2 (dotted diamonds) and FAGS2 (dotted circles). Comparing FAGS2 and CDCC2 (both with $U_{pA}$ at $E_p$ in the entrance channel) one concludes that the discrepancy is almost independent of the ambiguity in $U_{pA}$. As in \cite{nunes11b} we do find
an increasing dependence on the choice of the optical potentials with beam energy. CDCC calculations should be independent
of the choice of the auxiliary interaction. This is true for the lowest energy, but a few percent discrepancy starts
to appear as the beam energy increases.

In Fig. \ref{c12tr}, we show $d \sigma/d \Omega$ for $^{12}$C(d, p)$^{13}$C reaction at $E_{\rm d}$ = 12  and 56 MeV. Just as in the case of $^{10}$Be, the CDCC predictions for (d,p)
on $^{12}$C at low energy provides a very good approximation to the Faddeev solution (FAGS1).
However, the disagreement at 56 MeV becomes
significant (around $20$ \%).  Whereas in $^{10}$Be there was a strong dependence of the transfer cross section on the choice of the energy at which the optical potentials are evaluated, for $^{12}$C, no such
dependence exists (compare FAGS1 and FAGS2 or CDCC and CDCC2) and therefore the disagreement is quantitatively robust.

In Fig.\ref{ca48tr} we present the angular distributions following (d,p) transfer to the ground state of $^{49}$Ca
at 56 MeV. Small but not negligible discrepancies are found in the shapes of $d \sigma/d \Omega$
between FAGS1 and CDCC, accompanied by a significant
dependence on the optical potential (FAGS2 and CDCC2), which makes the comparison ambiguous.
%@@@@@@@@@@@@@@@@@@@@@@@@@@@@@@@@ Table 3 @@@@@@@@@@@@@@@@@@@@@@@@@
\begin{center}
\begin{table}[t!]
{\small
\hfill{}
\vspace{10pt}
\begin{tabular}{|c|c|r|r|r|}
\hline
\hline
\vspace{0.1cm}
target & $E_{d}$ (MeV)  &   {\small $\Delta_{\rm A-C}$ (\%)}  & {\small $\Delta_{\rm F-C}$ (\%)} & {\small $\Delta_{\rm F-A}$ (\%)} \\
\hline
		    & $21.4$    & $-3$  & $3 \pm 1$    & $6 \pm 1$   \\
$^{10}$Be  	& $40.9$    & $-21$ & $-36\pm 19$  & $-19 \pm 19$  \\
		    & $71.0$    & $-9$  & $-53 \pm 47$ & $-48 \pm 47$  \\
\hline
$^{12}$C 	& $12.0$    & $8$   & $6 \pm 3$    & $-2 \pm 3$   \\
		    & $56.0$    & $13$  & $-21 \pm 1$  & $-30 \pm 1$   \\
\hline
$^{48}$Ca 	& $56.0$    & $-5$  &  $39\pm 14$  & $47 \pm 14$ \\
\hline
\hline
\end{tabular}
\hfill{}}
\caption{\label{error}Percentage differences between the (d,p)
cross sections at the peak of the angular distribution predicted by the various methods,
for the three targets here considered, as a function of beam energy $E_d$.}
\end{table}
\end{center}
%@@@@@@@@@@@@@@@@@@@@@@@@@@@@@@@@@@ Ends @@@@@@@@@@@@@@@@@@@@@@@@@@

We connect the present work with the comparative study \cite{nunes11b} between the finite-range Adiabatic wave
approximation (ADWA) method and Faddeev-AGS. For that purpose, we include
in  Figs. \ref{be10tr}, \ref{c12tr} and \ref{ca48tr}, the finite-range ADWA predictions (stars-short-dashed).
These correspond to exactly the same Hamiltonian as the CDCC calculations performed here,
and therefore differ from those in \cite{nunes11b}.
Encouragingly, for the reactions on $^{10}$Be, ADWA performs just as well
or even better than CDCC. Fig. \ref{be10tr} is further evidence
that,  for $l=0$ transfer, ADWA  provides a very good approximation to Faddeev-AGS for reactions around $10$ MeV/u, as concluded in \cite{nunes11b}.
For $^{12}$C and $^{48}$Ca, ADWA cross sections differ from the CDCC predictions by a few percent, usually
increasing the discrepancy with the FAGS1 results (as compared to CDCC).

Table \ref{error} provides a quantitative estimate for the discrepancy between the various methods here considered.
We provide percentage differences of the differential cross sections at the peak of the angular distributions:
i) $\Delta_{A-C}$ compares ADWA with the standard CDCC, relative to CDCC,
ii) $\Delta_{F-C}$ compares FAGS1 with CDCC, relative to CDCC, and iii) $\Delta_{F-A}$
compares FAGS1 with Adiabatic, relative to ADWA.
The error in $\Delta_{F-C}$ and $\Delta_{F-A}$ is estimated  from the percentage difference
obtained with FAGS2 versus FAGS1.

% ----------------------------------------------------------------------

\subsection{\label{bu}Deuteron Breakup}

In this section, we compare CDCC angular and energy distributions following the breakup of the deuteron
with the Faddeev-AGS counterparts.  We consider FAGS and FAGS1, i.e.,
without and with a  bound state in the nA subsystem.

Breakup calculations are computationally more demanding than transfer or elastic.
As mentioned in Sec.~\ref{fags-model-space}, the Faddeev-AGS breakup results obtained
with the  present technical implementation are not sufficiently accurate at forward angles
when the Coulomb force is included.
For this reason, the comparison of breakup cross sections is performed switching off the Coulomb interaction in both CDCC and Faddeev-AGS.
\footnote{Based on CDCC predictions, the inclusion of Coulomb can increase the cross section by up to a factor of 2.},

\begin{figure}[t!]
\begin{center}
\includegraphics[scale=0.36]{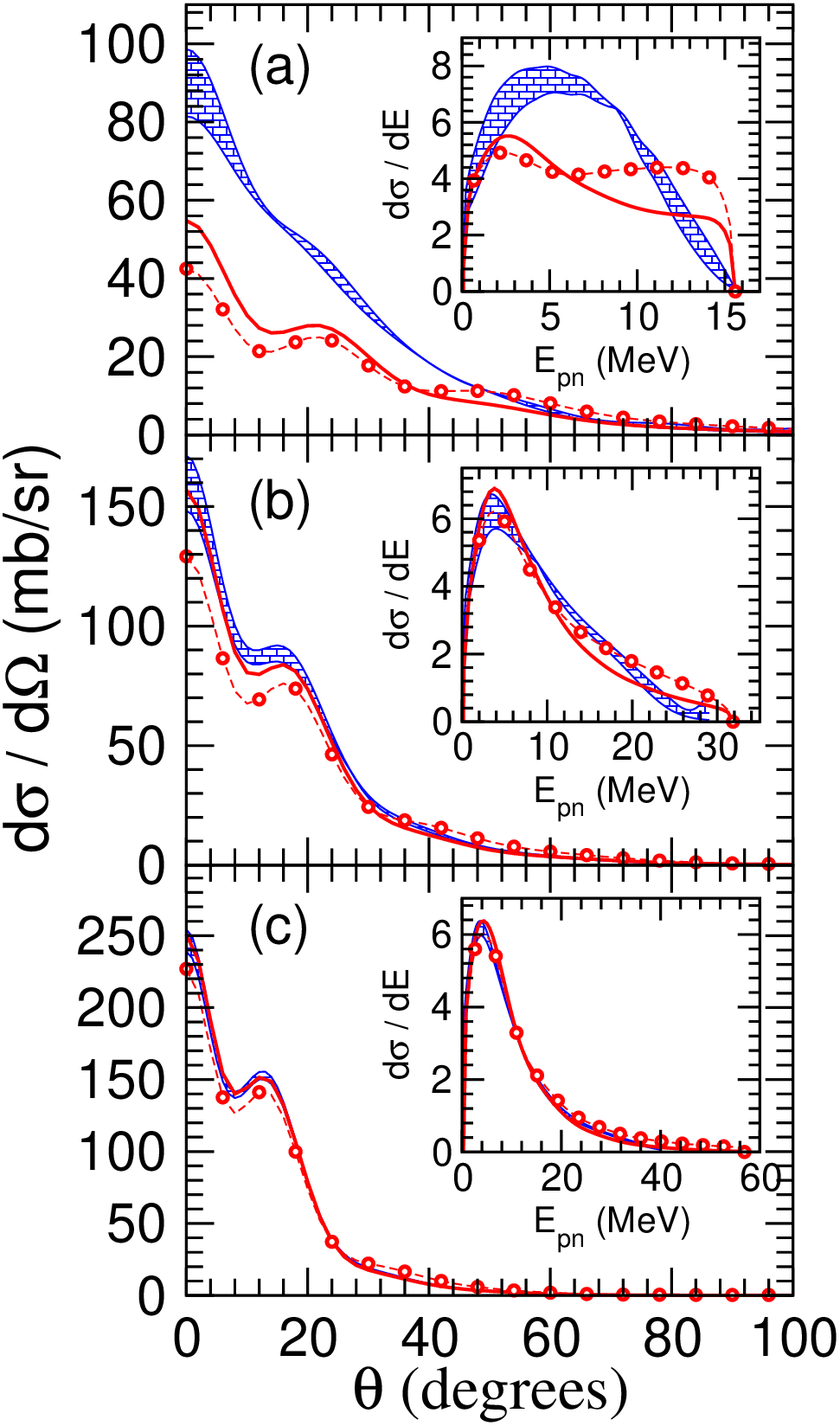}
\end{center}

\vspace{-1.2em}
\caption{\label{be10bu}Breakup distributions for $^{10}$Be (d, pn) $^{10}$Be reaction at: (a)
$E_{\rm d}$ = 21 MeV, (b) $E_{\rm d}$ = 40.9 MeV and (c)
$E_{\rm d}$ = 71 MeV. Results for CDCC (hashed bar), FAGS(solid) and FAGS1 (circles). }
\vspace{-1.2em}
\end{figure}

In its present implementation, CDCC breakup observables are reconstructed from
the asymptotic form of the three-body wavefunction. This implies that the asymptotics of the three-body wavefunction needs to be good not only when the neutron and proton are close to each other,
but also when they are far apart. This poses considerable  challenges
to our numerical methods. For all the CDCC calculations, we went to the numerical
limit of the methods, mostly dictated by the number of partial waves included in the deuteron continuum.
For the nuclear-only CDCC breakup calculations, $Q_{max}=l_{max}=6$ was our model space limit.
The error bands in the CDCC cross sections are due to the truncation of the model space, and have been
extrapolated from the differences found in the cross sections for $Q_{max}=l_{max}=6$ and $Q_{max}=l_{max}=4$.
Typically the calculations at higher energies have a smaller error bar because for those energies
the effect of these higher multipole couplings is smaller.

In Fig.\ref{be10bu} we present the results for the angular distribution as a function of the c.m. angle of the $pn$ system following the breakup on $^{10}$Be at the three energies of choice. The hashed band, the solid line and the circles correspond to CDCC, FAGS and FAGS1 predictions, respectively.
At the lowest energy,  we find that CDCC does not reproduce FAGS, even taking into account the error estimated by model space truncation.  At the higher energies, this discrepancy is removed.
The insets of Fig.\ref{be10bu} contain the corresponding energy distributions as functions of the proton-neutron
 relative energy  $E_{pn}$. Again, a very large discrepancy is found at $21.4$ MeV while fair agreement
between CDCC and FAGS is obtained at the higher energies.

\begin{figure}[t!]
\begin{center}
\includegraphics[scale=0.36]{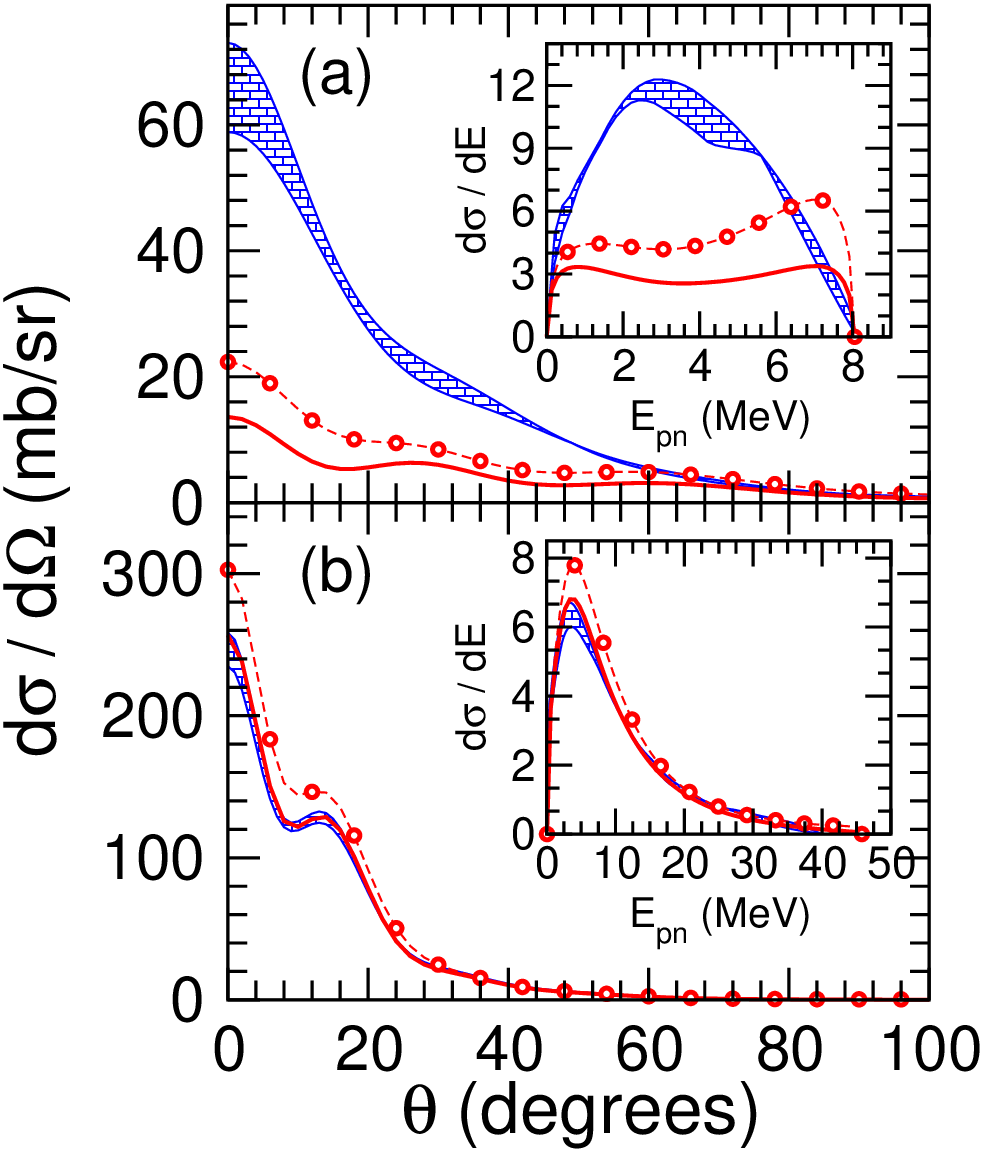}
\end{center}

\vspace{-1.2em}
\caption{\label{c12bu}Breakup  distributions  for $^{12}$C (d, pn) $^{12}$C reaction at: (a)
$E_{\rm d}$ = 12 MeV and (b) $E_{\rm d}$ = 56 MeV. Results for CDCC (hashed bar), FAGS(solid) and FAGS1 (circles). }
\begin{center}
\includegraphics[scale=0.36]{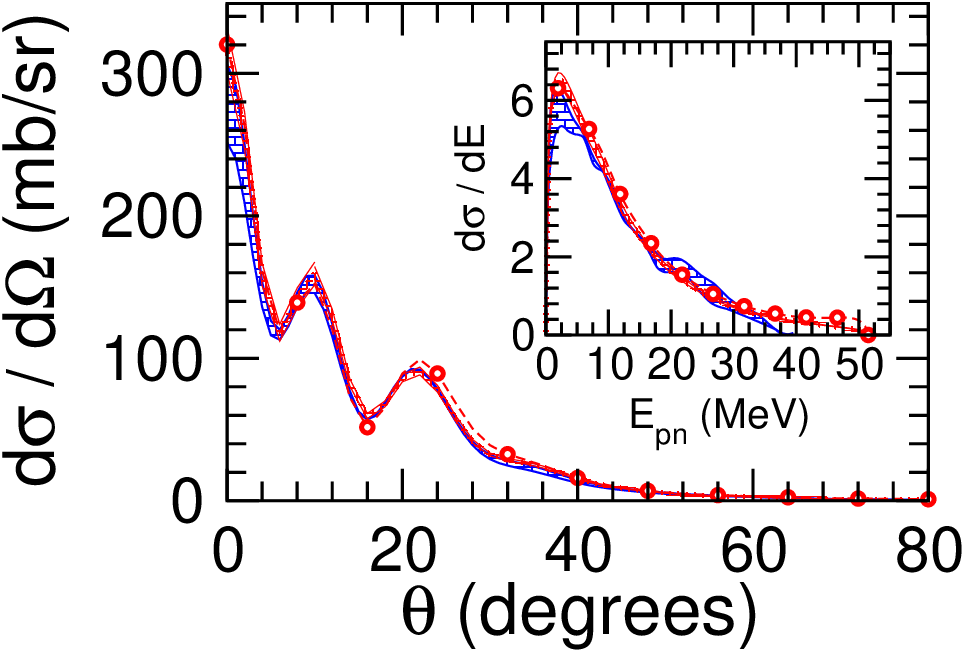}
\end{center}

\vspace{-1.2em}
\caption{\label{ca48bu}Breakup distributions  for $^{48}$Ca (d, pn) $^{48}$Ca reaction at $E_{\rm d}$ = 56 MeV. Results for CDCC (hashed bar), FAGS(solid) and FAGS1 (circles). }
\vspace{-1.2em}
\end{figure}

Similar conclusions can be drawn from the comparison of breakup angular and energy distributions for reactions on $^{12}$C (Fig.\ref{c12bu}).  Despite the large error bar in the CDCC predictions,
there is a striking mismatch between CDCC and FAGS in both magnitude and shape of the breakup cross sections
 at 12 MeV.
These discrepancies disappear at the higher energy.
Agreement is obtained between CDCC and FAGS for the breakup
of deuterons on  $^{48}$Ca at $56$ MeV, as shown in Fig.\ref{ca48bu}.

The effects of including the nA bound state in the transfer channel is shown with FAGS1 (dotted circles).
By comparing FAGS and FAGS1 we conclude that the effects of transfer are not negligible on breakup,
particularly at low energies.

Although for transfer we found excellent agreement for all reactions
at low energy ($\approx 10$ MeV/u), for breakup it is exactly at the low energy
that CDCC appears to break down. To better understand this, we have investigated
the relevant contributions to the breakup cross sections within FAGS.
In particular, we wanted to understand the relative importance of the three
Faddeev components, namely the deuteron channel, the proton channel and the neutron channel,
in evaluating the breakup observables. We suspected this was the key to understanding
the discrepancies between CDCC and FAGS, because CDCC is based on an expansion in
the deuteron channel alone.
Even though in the FAGS calculations
there are no bound states in the proton
and the neutron channels  ($U_{nA}$ and $U_{pA}$ contain an imaginary term),
the breakup is still distributed between all three Faddeev components.
We find that, for breakup at low beam energies, the proton and neutron Faddeev
components are equally important as the deuteron component, and there are destructive interferences
between the various components, at most values of $E_{pn}$, that reduce the cross section significantly.

At the limiting case where all the energy is in the proton-neutron relative motion,
the phase space factor goes as $\sim q_A \sim \sqrt{E_{pn}^{\rm max}-E_{pn}}$, where $q_A$ is the
target recoil momentum in the c.m. system. This is not
well reproduced when expanding in terms of a truncated deuteron basis, as in CDCC.
Given the coordinate choice, it is expected that the asymptotic behavior of the CDCC wavefunction is at its best when the proton and neutron are close to each other, but not in the opposite case.
The results presented in Figs. \ref{be10bu}, \ref{c12bu}, \ref{ca48bu} are a manifestation of this fact.
CDCC breaks down at the lower energy regime,  where we find broader angular and energy distributions
and there are larger
contributions to the breakup cross section from $E_{pn} \approx E_{pn}^{\rm max}$.

Our previous work \cite{nunes1} suggested that the discrepancies between CDCC and Faddeev in
breakup arrived as a consequence of the existence of a bound state in the rearrangement channel.
Results shown in Figs.\ref{be10bu} and \ref{c12bu} demonstrate this is  not always the case.
Even if there is no nA bound state, large discrepancies can occur at low energies.
The inclusion of a bound state enhances the disagreement, as shown by FAGS1 results in Figs. \ref{be10bu},\ref{c12bu}, and \ref{ca48bu}.

%%%%%%%%%%%%%%%%%%% CONCLUSIONS  %%%%%%%%%%%%%%%%%%%%%%%%%%%%%%
\section{Summary and Conclusions}

In this study we perform a systematic comparison between  CDCC and  Faddeev-AGS
for deuteron induced reactions. The CDCC formalism under study here corresponds to the standard method
introduced in \cite{cdcc}, and is based on a truncated expansion on the deuteron continuum.
The Faddeev-AGS method considered in this work corresponds to the solution of the AGS equations \cite{ags},
including Coulomb screening and renormalization as in \cite{deltuva05}. For all practical purposes, Faddeev-AGS
solutions  are considered exact and therefore this study should serve as a test on
the reliability of CDCC.

We focus on deuteron reactions on $^{10}$Be, $^{12}$C and $^{48}$Ca, including a wide range of beam energies.
We compute elastic scattering, transfer cross sections to the
ground state of the final system, as well as breakup observables. In CDCC, elastic scattering and breakup
cross sections are obtained directly from the S-matrix, while transfer is calculated replacing the exact
three-body wavefunction by the CDCC wavefunction in the exact post-form T-matrix.

Our CDCC/FAGS comparisons show no immediate correlation between elastic, transfer or breakup.
In other words, finding agreement for the elastic for a given target and beam energy does not imply
agreement in breakup or transfer.
Indeed, these processes are sensitive to different parts
of configuration space and therefore, only by looking at elastic, transfer and breakup simultaneously,
can the CDCC method be thoroughly tested.

Overall, and regardless of the beam energy, CDCC is able to provide a good approximation to FAGS
for elastic scattering. The inclusion of a neutron-nucleus bound state in the FAGS1 calculations
introduces small modifications mostly are backward angles. Only for  d$+^{12}$C at $12$ MeV
we found stronger discrepancies in the elastic angular distribution between CDCC and Faddeev-AGS.

The comparison of CDCC and Faddeev-AGS for transfer cross sections is consistent with the results
presented in \cite{nunes11b}. We found CDCC to be a very good approximation of FAGS1 at
reactions around 10 MeV/u, but not so good for larger beam energies.

As opposed to transfer, breakup observables predicted by CDCC are at its best
for the higher beam energies explored in this work. To reduce the technical challenges of
the problem, we ignore the Coulomb interaction in the breakup comparison. Also, we use exactly
the same Hamiltonian (CDCC and FAGS) to remove any ambiguity.  Taking into account the
estimated error due to the truncation of the model space in the CDCC calculations,
CDCC predictions for the breakup angular and energy distributions are in good
agreement with FAGS for all but the lowest energies considered.
For deuteron breakup on $^{10}$Be at 21.4 MeV and on $^{12}$C at 12 MeV, CDCC fails.
Strong contributions from the proton and neutron Faddeev components, not explicitly included
in the CDCC expansion, are present when the proton-neutron relative energies are large.
At low energy, the energy distribution is broad, the breakup to scattering states with
large proton-neutron relative energy is important and therefore CDCC does not
perform well. One possible solution to this shortcoming is to use the
CDCC wavefunction in a T-matrix that probes only short distances between the proton and neutron,
instead of its asymptotic form.

The present comparisons pose important constraints on the validity of CDCC when applied
to deuteron induced reactions. However, it is critical to realize that the number of reactions which
can be calculated with the present implementation of Faddeev-AGS is more limited.
Indeed, as the mass of the target increases, Coulomb effects become stronger requiring
larger screening radius and more partial waves; this renders the solution of the AGS equations
in the partial-wave representation impossible. In addition, the Pade summation technique
used to solve the AGS equations converges slower at low energies (10 MeV/u), where couplings are
strong, and it is not possible to obtain converged solutions even for mass $A=48$.
It would be desirable to extend present techniques used in Faddeev-AGS or to
develop new methods which can overcome these difficulties.

Finally, one should keep in mind that this work is based on a pure three-body formulation
of deuteron induced reactions. This assumes that the final bound state populated in the (d,p) reaction is
of pure single particle nature. A great challenge and important advance for the Faddeev-AGS formulation is
the inclusion of degrees of freedom of the target, which is necessary for a more realistic
description of the process.

% ----------------------------------------------------------------------
\medskip
%%%%%%%%%%%%%%%%%%%%%%%% ACKNOWLEDGEMENTS  %%%%%%%%%%%%%%%%%%%%%%%%%%%%%
We thank Ron Johnson and the members of the TORUS collaboration for important discussions
on this work. We are in debt to Ian Thompson for his invaluable support
with {\sc fresco}.
The work of N.J.U. was supported by the Department of Energy topical collaboration grant DE-SC0004087.
The work of A.D. was partially supported by the FCT grant PTDC/FIS/65736/2006.
The work of  F.M.N. was partially supported by the National Science
Foundation grant PHY-0800026 and the Department of Energy through grant
DE-FG52-08NA28552.
% ----------------------------------------------------------------------

%%%%%%%%%%%%%%%%%%%%%%%%%%%%%% REFERENCES  %%%%%%%%%%%%%%%%%%%%%%%%%%%%%

% ----------------------------------------------------------------------
\end{document}